\documentclass[12pt, onecolumn, draftcls]{IEEEtran}
\usepackage[latin1]{inputenc}
\usepackage[english]{babel}
\usepackage[T1]{fontenc}
\usepackage{array}
\usepackage{graphicx}
\usepackage[caption=false,font=footnotesize]{subfig}
\usepackage{fixltx2e}
\usepackage{cite}
\usepackage{algorithm}
\usepackage{algorithmic}
\usepackage{subfig}
\usepackage{wrapfig}	
\usepackage[cmex10]{amsmath}
\usepackage{amsfonts,amssymb}
\usepackage{bm}
\usepackage{booktabs}
\usepackage{color}
\usepackage{epstopdf}
\usepackage{float}
\usepackage{multirow}
\usepackage{multicol}
\usepackage{soul}
\usepackage{url}
\usepackage{indentfirst}
\usepackage{tikz}
\definecolor{gold}{rgb}{0.85,.66,0}
\definecolor{green}{rgb}{0.01,0.5.0.01}
\definecolor{cian}{rgb}{.02,.7,.95}

\newcommand{\colk}{\textcolor{black}}

\usepackage{xspace,colortbl}

\begin{document}

\title{\colk{Massive MIMO Pilot Assignment Optimization based on Total Capacity}}
\author{Jose~Carlos~Marinello~Filho, Cristiano~Magalhães~Panazio, Taufik~Abrão\thanks{T. Abrão and J. Marinello are with Department of Electrical Engineering, State University of Londrina, PR, Brazil. C. Panazio is with Laboratory of Communications and Signals, Polytechnic School of the University of Sao Paulo, SP, Brazil. E-mail: taufik@uel.br; \,\, zecarlos.ee@gmail.com; \,\, cpanazio@lcs.poli.usp.br}
\vspace{-3mm}}
\date{\today}

\maketitle

\begin{abstract}
\colk{We investigate the effects of} pilot assignment (PA) in multi-cell massive multiple-input multiple-output (Ma-MIMO) systems. When deploying a large number of antennas at base station (BS), and linear detection/precoding algorithms, the system performance in both uplink (UL) and downlink (DL) is mainly limited by pilot contamination. This interference is proper of each pilot, and thus system performance can be improved by suitably assigning the pilot sequences to the users within the cell, according to the desired metric. We show in this paper that UL and DL performances constitute conflicting metrics, in such a way that \colk{one cannot achieve the best performance in UL and DL with a single pilot assignment configuration. Thus, we propose an alternative metric, namely total capacity}, aiming to simultaneously achieve a suitable performance in both links. Since the PA problem is combinatorial, and the search space grows with the number of pilots in a factorial fashion, we also propose a low complexity suboptimal algorithm that achieves promising capacity performance avoiding the exhaustive search. Besides, \colk{the combination of our proposed PA schemes with an efficient power control algorithm unveils} the great potential of the proposed techniques in providing improved performance for a higher number of users. Our numerical results demonstrate that with 64 BS antennas serving 10 users, our proposed method can assure a 95\%-likely rate of 4.2 Mbps for both DL and UL, and a symmetric 95\%-likely rate of 1.4 Mbps when serving 32 users.

\end{abstract}

\section{Introduction}\label{sec:intro}%
Pilot contamination is the most salient impairment of massive multiple-input multiple-output (Ma-MIMO) systems \cite{Marzetta10}, \cite{Marzetta16}. Due to this issue, it can be shown that interference seen by each user is dependent on the pilot sequence he is employing. Thus, one can improve system performance by suitably assigning the pilots to the users in the cell, as discussed in \cite{SPA_Zhu15} and \cite{Marinello2016}. The pilot assignment (PA) problem was solved in \cite{SPA_Zhu15} from the uplink (UL) perspective, and a low-complexity near-optimal solution was proposed. Such simple and very efficient solution was possible due to the simple dependence between performance and pilot assignment seen in UL. Pilot contamination interference in UL is due to the signals transmitted from users in adjacent cells sharing a given pilot sequence that reaches certain base station (BS). The user that will experience such interference is the one to which this pilot sequence is assigned in the cell. Under a max-min optimization perspective, the solution proposed in \cite{SPA_Zhu15} simply assigns the worst pilot sequences in a given instant, in terms of pilot contamination arising from adjacent cells, to the best located users, \emph{i.e.}, those with the higher long-term fading coefficients. 

In \cite{Marinello2016}, the PA problem was investigated from the downlink (DL) perspective, under several metrics. However, as the dependence of DL signal-to-interference-plus-noise ratio (SINR) with the pilot assignment occurs in a more complex way than in UL, the problem was solved only via exhaustive search. Different than UL, pilot contamination interference in DL is due to the signals transmitted from neighbouring BS's to the users at their respective cells sharing certain pilot sequence, but that inadvertently reaches the user in the considered cell employing this pilot. If this user changes its pilot sequence, part of the pilot contamination reaching him remains unchanged, since the distance between the user and adjacent cells as well as the long-term fading coefficients remain the same. The part of interference that changes is due to the power normalization factors, that retain some dependence with the UL pilot transmission stage, as better clarified in this manuscript, which also proposes low-complexity DL pilot assignment schemes.

A greedy PA scheme is proposed for optimizing the DL performance of cell-free Ma-MIMO systems in \cite{Ngo17}. This new concept of Ma-MIMO system comprises a very large number of distributed single-antenna access points simultaneously serving a much smaller number of users. Although having considerably improved performance with respect to conventional collocated Ma-MIMO systems due to the diversity of long-term fading coefficients, the system implementation cost is significantly higher due to the complex backhaul network and the very large number of access points to be installed. Besides, the proposed PA scheme aims to improve just the DL performance. As shown in this paper, UL and DL optimization metrics constitute conflicting objectives, since if one scheme optimizes system DL performance, the UL performance will be close to that \colk{attained} with random assignment, and vice-versa. Thus, our main purpose in this paper is to design a PA methodology able to jointly optimize DL and UL perspectives.

\colk{Another promising PA technique is proposed in \cite{Zhu17} to improve UL performance, in which a weighted graph-coloring-based pilot decontamination scheme is developed. The solution consists in constructing an edge-weighted interference graph aiming to depict the potential pilot contamination between users, whereby two users in different cells are connected by a weighted edge, indicating the interference strength when they reuse the same pilot. Then, inspired by classical graph coloring algorithms, the proposed solution denotes each color as a pilot and each vertex as a user in the interference graph, which is able to improve performance by assigning different pilots to connected users with a large weight in a greedy way. On the other hand, an efficient PA technique for improving DL performance of wideband massive MIMO systems is proposed in \cite{Chen16} by exploiting channel sparsity. Authors employ Karhunen-LoÃ©ve Transform and Discrete Fourier Transform to capture
the hidden sparsity of the channel, and find that the subspaces of the desired and interference channels are approximately orthogonal when the channels are represented with the aid of DFT basis. Then, a pilot assignment policy is designed to help identify the subspace of the desired channel, and a desired channel subspace aware least square channel estimator is derived to remove pilot contamination.}

Several pilot assignment schemes have been proposed recently based on the location-aware approach \cite{Zhao16}, \cite{Wang16}. Authors have proposed some filtering techniques which are able to significantly remove pilot contamination from channel estimates if the training signals transmitted from users of different cells sharing a given pilot reach the considered BS with different angle-of-arrivals (AOAs). For this sake, the location-aware PA technique assigns the pilots for users in the considered cell aiming to ensure that users utilizing the same pilot have distinguishable AOAs. However, all these schemes rely on the existence assumption of a line-of-sight component between BS and users, and of small channel angle spread of the multi-path components observed at the BS, typically observed in rural and sub-urban areas, or if the BS is much higher than the surrounded structures with few scatters around \cite{Wang16}. Since this is not always the case, we investigate in this paper PA schemes to be deployed in a less restrictive \colk{scenarios}.

Power allocation is another efficient way of improving the performance of wireless communication networks. For the multicell massive MIMO scenario, the problem of power allocation was investigated in \cite{Zhang15} under the perspective of maximizing the sum rate per cell. It was shown that the proposed method achieves substantial gains over the equal power allocation policy. However, for a practical system, maximizing the sum rate per cell is not the most suitable objective, since the performance of some users (tipically those at the edge) may be severely penalized in order to provide very increased rates for another ones. \colk{In \cite{Zarei17}, a max-min power allocation policy is proposed in conjunction with a multicell-aware regularized zero-forcing precoding to improve fairness in the massive MIMO DL. The proposed technique achieves substantially higher network-wide minimum rates than conventional techniques, but is applicable only for DL.} On the other hand, in \cite{Debbah16}, a power allocation technique able to provide the same performance for all served users in DL and UL is proposed. The UL power allocation policy allocates for each user a power proportional to a desired received signal-to-noise ratio (SNR) at the BS divided by its long-term fading coefficient. Since shadowing have not been assumed in that work, one has just to ensure that this desired SNR level allows the cell boundary user to transmit without exceeding the maximum transmit power of its device. Clearly, in a more realistic scenario, a severely shadowed user near the cell edge may not be able to transmit with such power. A fairer optimization metric is adopted by the power control algorithm proposed in \cite{Rasti11} for a code division multiple access network, which consists of providing a target performance for the majority of the users in the network. Since such approach is very suitable in the context of Ma-MIMO systems, we have adopted this distributed power control algorithm herein.

In this paper, we address the PA problem from jointly UL and DL perspectives. Since the performance bottleneck of Ma-MIMO systems is due to the edge users suffering with severe pilot contamination \cite{Marzetta10}, we consider the objective of providing a target performance for the majority of the users. For this sake, the pilot allocation metric of maximizing the minimum SINR is the most suitable, and our proposed scheme amalgamates such metric with the power control algorithm of \cite{Rasti11}, by only knowing the power and the long-term fading coefficients of users in adjacent cells. We first propose a low-complexity suboptimal solution of the MaxminSINR DL PA problem of \cite{Marinello2016}. Then, we show that the optimization under DL and UL perspectives are conflicting, since the optimal performance in both directions cannot be simultaneously achieved with any pilot configuration. We thus propose an alternative PA procedure from the definition of the overall \colk{system} capacity, as the sum of DL and UL capacities. Our numerical results demonstrate that, under this metric, significant gains can be achieved in DL and UL concomitantly. Besides, a low complexity suboptimal approach for solving the joint UL-DL PA problem is discussed. Finally, we adapt the power control algorithm of \cite{Rasti11} to the Ma-MIMO scenario with finite number of BS antennas, and show that much more significant gains can be attained by the PA schemes when combined with an efficient power control algorithm.

\colk{Our main contributions can be summarized as follows: {\bf i}) different than \cite{Marinello2016} that solved the DL PA problem only via exhaustive search, we propose a near optimal low-complexity method for solving it; {\bf ii}) different than \cite{SPA_Zhu15}, \cite{Marinello2016}, \cite{Ngo17}, \cite{Zhu17}, \cite{Chen16}, we investigate the PA optimization problem from both UL and DL perspectives, showing that a single pilot assignment configuration cannot lead to the optimal performance in both links at the same time; {\bf iii}) we then formulate a novel PA optimization metric, namely the total capacity, which is able to achieve good performances in both links concomitantly;  {\bf iv}) a near optimal low-complexity solution for the total capacity PA problem is also proposed;  {\bf v}) we then combine the investigated PA techniques with an efficient power control algorithm, showing that even more impressive performances can be achieved.}

The rest of this paper is organized as follows. The system model is described in Section \ref{sec:model}, while the proposed pilot assignment optimization schemes are discussed in Section \ref{sec:optimization}. The adopted power control algorithm are described in Section \ref{sec:PowC}. Illustrative numerical results are explored in Section \ref{sec:results}. Final remarks and conclusions are  offered in Section \ref{sec:concl}.

\section{System Model}\label{sec:model}
We consider a similar system model of \cite{SPA_Zhu15}, \cite{Marinello2016}, which is composed of $L$ hexagonal cells, each one equipped with a $N$ antennas BS serving $K$ single-antenna users. Time division duplex (TDD) is assumed, thus reciprocity holds, and channel state information (CSI) is acquired by UL training sequences transmission. The $N \times 1$ channel vector between the BS of $i$-th cell and the $k'$-th user of $j$-th cell is denoted by ${\bf g}_{ik'j} = \sqrt{\beta_{ik'j}} {\bf h}_{ik'j}$, in which $\beta_{ik'j}$ denotes the long-term fading coefficient, comprising path loss and log-normal shadowing, ${\bf h}_{ik'j} \sim \mathcal{CN}({\bf 0}_N, {\bf I}_N)$ is the short-term fading channel vector, while ${\bf 0}_N$ is a null column vector of size $N \times 1$, and ${\bf I}_N$ is the identity matrix of size $N$.

The pilot \colk{sequences'} set is ${\bm \Psi} = [{\bm \psi}_1  \,\, {\bm \psi}_2 \, \ldots \, {\bm \psi}_K] \in \mathbb{C}^{K\times K}$, \colk{and the sequence length is also $K$}. This set of sequences is orthogonal, and thus ${\bm \Psi}^H {\bm \Psi} = {\bf I}_K$ holds, being $\{\cdot\}^H$ the Hermitian operator. If for the $k'$-th user is assigned the sequence ${\bm \psi}_k' = [\psi_{k' 1} \,\, \psi_{k' 2} \ldots \psi_{k' K}]^{T}$, being $\{\cdot\}^T$ the transpose operator, the received signal at the $i$-th BS during the training stage is
\begin{equation}\label{eq:rx_pilots}
{\bf Y}^{\rm p}_{i} = \sqrt{\rho^{\rm p}} \sum_{j = 1}^{L} \sum_{k = 1}^{K} {\bf g}_{ikj} {\bm \psi}^H_k + {\bf N}^{\rm p}_i,
\end{equation}
in which $\rho^{\rm p}$ is the UL pilot transmit power, ${\bf N}^{\rm p}_i \in \mathbb{C}^{N \times K}$ is the additive white Gaussian noise (AWGN) matrix with i.i.d. elements following a complex normal distribution with zero mean and variance $\sigma^2_n$. Note that we have assumed an uniform power allocation policy in the UL training stage. As discussed in \cite{Fernandes13}, any effect of pilot power allocation can be alternatively achieved through transmit power allocation and constant pilot powers, and thus no loss is incurred in assigning a constant pilot power $\rho^{\rm p}$ to all users. The $i$-th BS then estimates the $k'$-th user CSI by correlating the received signal ${\bf Y}^{\rm p}_{i}$ with ${\bm \psi}_k'$
\begin{equation}\label{eq:rx_correlator}
\widehat{{\bf g}}_{ik'} = \frac{1}{\sqrt{\rho^{\rm p}}} {\bf Y}^{\rm p}_{i} {\bm \psi}_{k'} = \sum_{j = 1}^{L} {\bf g}_{ik'j}  + {\bf v}_{ik'},
\end{equation}
where ${\bf v}_{ik'} = \frac{1}{\sqrt{\rho^{\rm p}}}{\bf N}^{\rm p}_i {\bm \psi}_{k'} \sim \mathcal{CN}({\bf 0}_N, \frac{\sigma^2_n}{\rho^{\rm p}} {\bf I}_N)$ is an equivalent noise vector. The pilot contamination effect can be clearly seen in the previous expression. By acquiring such \colk{CSI} estimates, the BS is able to perform linear detection in the UL and linear precoding in the DL, deploying maximum ratio combining (MRC) and maximum ratio transmission (MRT), respectively. \colk{It is important to note that only BS has CSI estimates, obtained directly from UL pilot transmissions in \eqref{eq:rx_correlator}, and thus no feedback channel is required.} During UL data transmission, the $i$-th BS receives the signal
\begin{equation}\label{eq:rx_data_u}
{\bf y}^{\rm u}_{i} = \sum_{j = 1}^{L} \sum_{k = 1}^{K} \sqrt{\rho^{\rm u}_{kj}} {\bf g}_{ikj} x^{\rm u}_{k j} + {\bf n}^{\rm u}_i,
\end{equation}
in which $\rho^{\rm u}_{kj}$ and $x^{\rm u}_{k j}$ are the UL data transmit power and the data symbol, respectively, from the $k$-th user of the $j$-th cell, and ${\bf n}^{\rm u}_i \sim \mathcal{CN}({\bf 0}_N, \sigma^2_n {\bf I}_N)$ is the $N\times 1$ AWGN sample vector. This BS then \colk{estimates} the transmitted symbol as 
\begin{equation}\label{eq:est_data_u}
\widehat{x}^{\rm u}_{k' i} = \widehat{{\bf g}}_{ik'}^H {\bf y}^{\rm u}_{i}.
\end{equation}

Similarly, during a DL data transmission, the $k'$-th user of the $j$-th cell receives the signal
\begin{equation}\label{eq:rx_data_d}
y^{\rm d}_{k' j} = \sum_{i = 1}^{L} \sum_{k = 1}^{K} \sqrt{\rho^{\rm d}_{i k}} {\bf g}_{ik'j}^T {\bf p}_{i k} x^{\rm d}_{i k} + n^{\rm d}_{k' j},
\end{equation}
where $\rho^{\rm d}_{i k}$ and $x^{\rm d}_{i k}$ is the DL data transmit power and the DL data transmitted from the $i$-th BS to his $k$-th user, and $n^{\rm d}_{k' j} \sim \mathcal{CN}(0, \sigma^2_n)$ is an AWGN sample. Besides, ${\bf p}_{i k}$ is the beamforming vector that the $i$-th BS computes to precode its $k$-th user data. Deploying MRT, this vector is defined as \cite{Fernandes13}:
\begin{equation}\label{eq:MF_prec}
{\bf p}_{i k} = \frac{{\bf \widehat{g}}_{i k}^*}{||{\bf \widehat{g}}_{i k}||} = \frac{{\bf \widehat{g}}_{i k}^*}{\alpha_{i k} \sqrt{N}},
\end{equation}
in which the scalar $\alpha_{i k} = \frac{||{\bf \widehat{g}}_{i k}||}{\sqrt{N}}$ is a normalization factor necessary to guarantee that $||\sqrt{\rho^{\rm d}_{i k}} {\bf p}_{i k} x^{\rm d}_{i k}||^2 = \rho^{\rm d}_{i k}$.

Following the analysis in \cite{Marzetta16}, it can be shown from \eqref{eq:est_data_u} and \eqref{eq:rx_data_d} that the UL and DL SINR performance of Ma-MIMO systems with MRC and MRT are given, respectively, by
\small
\begin{eqnarray}\label{eq:UL_SINR}
\varsigma^{\rm u}_{k' i} = \frac{\rho^{\rm u}_{k'i} \beta^2_{i k' i}}{\sum^{L}_{\substack{l = 1 \\ l \neq i}} \rho^{\rm u}_{k'l} \beta^2_{i k' l} + \frac{\alpha^2_{i k'}}{\colk{N}} \left(\sum_{l=1}^{L} \sum_{k=1}^{K} \rho^{\rm u}_{k l} \beta_{i k l} + \sigma^2_n \right)},
\end{eqnarray}
\begin{eqnarray}\label{eq:DL_SINR}
\varsigma^{\rm d}_{k' j} = \frac{\rho^{\rm d}_{j k'} \beta^2_{j k' j}/\alpha^2_{j k'}}{\sum^{L}_{\substack{l = 1 \\ l \neq j}} \rho^{\rm d}_{l k'} \beta^2_{l k' j}/\alpha^2_{l k'} + \frac{1}{\colk{N}} \left(\sum_{l=1}^{L} \beta_{l k' j} \sum_{k=1}^{K} \rho^{\rm d}_{l k} + \sigma^2_n \right)},
\end{eqnarray}
\normalsize
in which $\alpha^2_{j k} = \sum_{\ell = 1}^{L} \beta_{j k \ell} + \frac{\sigma_n^2}{\rho^{\rm p}}$. It is straightforward to see from \colk{\eqref{eq:UL_SINR}} and \colk{\eqref{eq:DL_SINR}} that the asymptotic UL and DL SINR ($N \to \infty$) converge, respectively, to
\begin{eqnarray}\label{eq:ULDL_SINR}
\varsigma^{{\rm u}\colk{\infty}}_{k' i} = \frac{\rho^{\rm u}_{k'i} \beta^2_{i k' i}}{\sum^{L}_{\substack{l = 1 \\ l \neq i}} \rho^{\rm u}_{k'l} \beta^2_{i k' l}}, \quad \quad \varsigma^{{\rm d}\colk{\infty}}_{k' j} = \frac{\rho^{\rm d}_{j k'} \beta^2_{j k' j}/\alpha^2_{j k'}}{\sum^{L}_{\substack{l = 1 \\ l \neq j}} \rho^{\rm d}_{l k'} \beta^2_{l k' j}/\alpha^2_{l k'}}.
\end{eqnarray}

Although UL and DL SINR \eqref{eq:ULDL_SINR} are similar-looking expressions, they have different statistical characteristics as discussed in \cite{Marzetta10}. While interference in UL is irradiated from users in neighboring cells using the same $k$-th pilot sequence to the $i$-th BS, interference in DL is irradiated from neighboring BS's to the user in $j$-th cell employing the $k$-th pilot sequence. While in UL the receiver is fixed and the multiple transmitters are moving, in DL multiple fixed transmitters communicate with a mobile receiver. Besides, different behaviors will be seen in each direction with respect to pilot assignment, as explained in the following.

\colk{The performance of the Ma-MIMO system can be improved both via pilot assignment and by power control. The optimal performance is achieved solving a joint optimization problem of increased complexity due to the mixed variables' types: the pilot assignment is of discrete nature, while power control is of continuous nature. In this paper, we adopt a decoupled approach, in which the pilot assignment is performed assuming uniform power allocation, followed by a power control algorithm evaluated from the optimized pilot distribution.}

\section{Pilot Assignment Optimization}\label{sec:optimization}
In this section, \colk{the Ma-MIMO system performance is improved} by suitably assigning the pilot sequences to the users \cite{SPA_Zhu15}, \cite{Marinello2016}. We assume a decentralized allocation procedure, in which PA is performed sequentially by each cell, regarding its covered users. The convergence of this procedure {after few PA optimization rounds} is discussed in \cite{Marinello2016}, and shown by numerical results in \cite{SPA_Zhu15}\footnote{{With the aid of game theory in \cite{Marinello2016}[Sec. 4.1], it is shown that the PA problem can be seen as a restricted potential game, in which each cell is a player that chooses its strategy following a selfish best response dynamics. Therefore, the convergence of the game to a Nash equilibrium is guaranteed. Moreover, in \cite{SPA_Zhu15}[Fig. 3(c)], it is noted that the number of optimization rounds to PA convergence among cells decreases as long as the number of antennas at BS increases. For instance, three rounds are sufficient for convergence under 128 BS antennas.}}. Besides, in this pilot assignment stage, \colk{it is assumed the knowledge of users' power distribution in the cells, and the focus is} at the asymptotic performance expressions in \eqref{eq:ULDL_SINR} \colk{for simplicity}. Then, in Section \ref{sec:PowC}, we evaluate the power control algorithm based on the pilot assignment obtained here, and demonstrate the performance improvements of our proposed schemes for finite number of antennas in Section \ref{sec:results}. SINR expressions in \eqref{eq:ULDL_SINR} were obtained assuming that the $k'$-th pilot sequence\footnote{\colk{With ``$k'$-th pilot sequence'', we refer to the pilot of index $k'$ in the set of available pilot sequences.}} is assigned to the $k'$-th user, \emph{i.e.}, under a random strategy. However, if we assume that the $p$-th pilot is assigned to the $c_p$-th user in the $i$-th cell, the UL SINR can be rewritten as
\begin{eqnarray}\label{eq:UL_SINR_PA}
\varsigma^{{\rm u}\colk{\infty}}_{c_p i} = \frac{\rho^{\rm u}_{c_p i} \beta^2_{i c_p i}}{\sum^{L}_{\substack{j = 1 \\ j \neq i}} \rho^{\rm u}_{p j} \beta^2_{i p j}},
\end{eqnarray}
that is the UL SINR of the user in the $i$-th cell employing the $p$-th pilot, namely the $c_p$-th user. \colk{The fixed uplink power distribution $\rho^{\rm u}_{p j}$, for $p=1,\ldots K$ and $j=1,\ldots L$, is known for the $i$-th cell.} The performance bottleneck in Ma-MIMO systems is due to users with severe pilot contamination \cite{Marzetta10}. Thus, a fair objective is to maximize the minimum UL SINR among the users, as the following optimization problem
\begin{equation}\label{eq:MaxminSINR_UL_PA}
\mathcal{PA}_{\rm u}: \max_{l} \min_{p} \quad \varsigma^{{\rm u}\colk{\infty}}_{c_{l p} i},
\end{equation}
\normalsize
in which $c_{l p}$ is the $l,p$-th element of matrix ${\bf C} \in \mathbb{N}^{K! \times K}$, which contains every possible pilot assignment that a given cell can adopt. $c_{l p}$ means that in the $l$-th PA combination, the $p$-th pilot is assigned to the $c_{l p}$-th user. \colk{According to \eqref{eq:MaxminSINR_UL_PA}, the obtained solution is the pilot assignment configuration in which the UL SINR of the user in the worst conditions is the highest possible. In terms of fairness and when aiming to provide a good and homogeneous performance to the users, this approach is much more effective than simply improving the average performance, as will be shown in Section \ref{sec:perf_comp}.} Obviously, the complexity of solving \eqref{eq:MaxminSINR_UL_PA} via an exhaustive search grows with $K$ in a factorial fashion. A viable and heuristic solution is proposed in \cite{SPA_Zhu15}, where the pilot sequences with the worst interference levels are assigned to the users with the highest long-term fading coefficients. Such {\it greedy method} achieves a near-optimal solution with reduced complexity.

On the other hand, the DL SINR in \eqref{eq:ULDL_SINR} can also be improved via pilot assignment. Assuming that the $p$-th pilot is assigned to the $c_p$-th user in the $j$-th cell, \colk{and that the downlink power distribution is known for the $j$-th cell}, the DL SINR becomes
\begin{eqnarray}\label{eq:DL_SINR_PA}
\varsigma^{{\rm d}\colk{\infty}}_{c_p j} = \frac{\rho^{\rm d}_{j c_{p}} \beta^2_{j c_{p} j}}{\beta_{j c_{p} j} + \vartheta_{j p (j)}} \, \frac{1}{\sum^{L}_{\substack{i = 1 \\ i \neq j}} \frac{\rho^{\rm d}_{i c_{p}} \beta^2_{i c_{p} j}}{\beta_{i c_{p} j} + \vartheta_{i p (j)}}},
\end{eqnarray}
\normalsize
where $\vartheta_{i p (j)} = \sum_{\substack{\ell = 1 \\ \ell \neq j}}^{L} \beta_{i p \ell} + \frac{\sigma_n^2}{\rho^{\rm p}}$ does not depend on for what user the $p$-th pilot will be assigned. Thus, an alternative PA optimization procedure for DL is proposed as:
\begin{eqnarray}\label{eq:MaxminSINR_DL_PA}
\mathcal{PA}_{\rm d}&:& \max_{l} \min_{p} \quad \varsigma^{{\rm d}\colk{\infty}}_{c_{l p} j},
\end{eqnarray}
\normalsize
\colk{whose solution, similarly as \eqref{eq:MaxminSINR_UL_PA}, is the pilot assignment configuration in which the DL SINR of the user in the worst conditions is the highest possible.}

\begin{figure*}[!htbp]
\centering
\includegraphics[width=1\columnwidth]{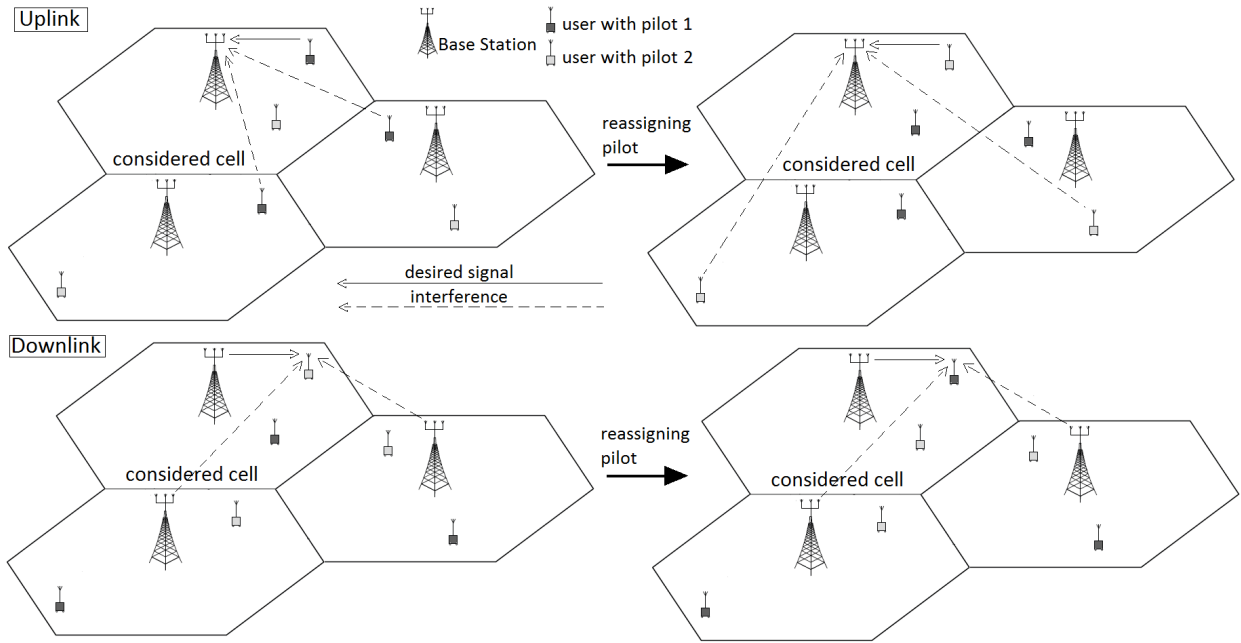}
\vspace{-8mm}
\caption{Effect of reassigning pilot sequences in the consired cell, in UL (up) and DL (down). Although it appears from the Figure that reassigning pilots has no effect in DL, the previous UL pilot transmission stage results in some variations on the signals' strength.}
\label{fig:Pilot_Assign}
\end{figure*}

One can see that the dependence of the DL SINR with the assignment of pilots to users is not so simple as it is in the UL scenario. However, communication systems are usually asymmetric, and DL data transmission might be predominant in comparison with UL data transmission. Therefore, solving the optimization problem in \eqref{eq:MaxminSINR_DL_PA} in a low-complexity way is also quite appealing in practice. The main difficulty is that the interference in DL for the $p$-th pilot (denominator of the second term in \eqref{eq:DL_SINR_PA}) is dependent on what user this sequence is assigned, and thus a greedy method cannot be applied as in \cite{SPA_Zhu15}. If a user changes its pilot sequence, the DL interference due to pilot contamination reaching him is still mainly dominated by his long-term fading coefficients with respect to neighboring BS's, which do not change, as illustrated in Figure \ref{fig:Pilot_Assign}. The exception is due to the DL normalization factors, that retain some dependence with the assigned pilot sequence from the uplink training channel estimation. 

\subsection{\colk{Joint UL-DL Pilot Assignment}}
One can note that the optimization problems of \eqref{eq:MaxminSINR_UL_PA} and \eqref{eq:MaxminSINR_DL_PA} are conflicting, as numerically demonstrated in the next Section. If the UL performance metric is optimized, the DL performance will be close to that attained with random assignment policy, and vice-versa. Thus, we elaborate our analysis aiming to jointly achieve the best possible performance in UL and DL, concurrently. One approach could be adopting multiobjective optimization. In this paper, we prefer a simpler strategy. We assume in the adopted TDD scheme a channel coherence block of $\mathcal{S}$ symbols, of which $K$ are devoted to UL pilot transmission, $\xi^{\rm u} \left( \frac{\mathcal{S} - K}{\mathcal{S}} \right)$ and $\xi^{\rm d} \left( \frac{\mathcal{S} - K}{\mathcal{S}} \right)$ for UL and DL data transmission, respectively, in which the factors $\xi^{\rm u}$ and $\xi^{\rm d}$ $\in (0,1)$, $\xi^{\rm u}+\xi^{\rm d} = 1$, {and both $\xi^{\rm u} \left( \mathcal{S} - K \right)$ and $\xi^{\rm d} \left( \mathcal{S} - K \right)$ are integers}. \colk{Note that the factor $\left( \frac{\mathcal{S} - K}{\mathcal{S}} \right)$ represents the pilot overhead, since for each coherence block of $\mathcal{S}$ symbols, $K$ are spent sending uplink pilots.} Then, for a system bandwidth $\textsc{bw}$, the total capacity (TC) of the user in the $i$-th cell for which the $p$-th pilot is assigned is defined as
\small
\begin{equation}\label{eq:cap_total}
\hspace{-.1mm}\mathcal{C}_{c_p i}^{\textsc{t}} = \textsc{bw}\left( \frac{\mathcal{S} - K}{\mathcal{S}} \right)\left[ \xi^{\rm u} \log_2\left(1 + \varsigma^{\rm u}_{c_p i}\right) + \xi^{\rm d} \log_2\left(1 + \varsigma^{\rm d}_{c_p i}\right) \right]\hspace{-1mm}
\end{equation}
\normalsize

Hence, an alternative optimization problem can be defined as
\begin{equation}\label{eq:MaxminCtt_PA}
\mathcal{PA}_{\textsc{tc}}: \max_{l} \min_{p} \quad \mathcal{C}_{c_{l p} i}^{\textsc{t}\colk{\infty}},
\end{equation}
\normalsize
in which $\mathcal{C}_{c_{l p} i}^{\textsc{t}\colk{\infty}}$ is obtained evaluating \eqref{eq:cap_total} from $\varsigma^{{\rm u}\colk{\infty}}_{c_p i}$ and $\varsigma^{{\rm d}\colk{\infty}}_{c_p i}$. \colk{In this case, the solution of \eqref{eq:MaxminCtt_PA} is the PA configuration in which the total capacity of the user in the worst conditions is the highest possible.}

\subsection{\colk{Heuristic Pilot Assignment Solutions}}

For solving the DL assignment problem avoiding the exhaustive search, we first create a cost matrix $\colk{\bf \Gamma}^{(i)} \in \mathbb{R}^{K \times K}$, whose $j,p$-th element is the DL SINR achieved by the $j$-th user in the $i$-th cell when the $p$-th pilot sequence is assigned to him, \colk{\it i.e.,} $\gamma^{(i)}_{jp} = \varsigma^{\rm d}_{c_p i}$ when $c_p = j$. Then, \emph{our proposed low-complexity suboptimal solution} consists of finding the pilot with which each user achieves its maximal DL SINR. The user with the lowest maximal DL SINR will have such pilot sequence assigned to him. This procedure is repeated, excluding the users and the pilots already assigned, until allocating every pilot sequence, \colk{as} described in Algorithm \ref{alg:PA_proced}.

In order to obtain a low-complexity form for solving \colk{the joint UL-DL PA in} \eqref{eq:MaxminCtt_PA} while achieving near-optimal performance, \colk{this same heuristic procedure can be applied. In this case,} we initialize the cost matrix $\colk{\bf \Gamma}^{(i)}$ (line \ref{genCM} in Alg. \ref{alg:PA_proced}) in an alternative way, such that $\gamma^{(i)}_{j p} = \mathcal{C}_{c_p i}^{\textsc{t}}$ when $c_p = j$, since the total capacity as defined in \eqref{eq:cap_total} is calculated as a weighted log function of both UL and DL SINR's. Then, we apply our proposed \colk{heuristic} method in a similar way as in Algorithm \ref{alg:PA_proced}.

\begin{algorithm}
\caption{\small{Proposed Pilot Assignment Procedure}}
\begin{flushleft}
Input: {$\beta_{j k l}$,  $\forall j,l =\nolinebreak 1,2,\ldots L$, $k= 1,2, \ldots K$, $\sigma_n^2$, $\rho_{\rm p}$.}
\end{flushleft}
\label{alg:PA_proced}
\begin{algorithmic}[1]
\STATE{Generate cost matrix $\colk{\bf \Gamma}^{(i)}$, of size $K\times K$;}\label{genCM}
\FOR{$j = 1, 2, \ldots, K$}
	\FOR{$p = 1, 2, \ldots, K$}
		\STATE{Evaluate $\delta_p = \max_{\ell = 1, \ldots, K} \gamma^{(i)}_{p,\ell}$;}
		\STATE{Evaluate $\eta_p = \arg \max_{\ell = 1, \ldots, K} \gamma^{(i)}_{p,\ell}$;} 
	\ENDFOR	
	\STATE{Evaluate $\phi = \arg \min_{p = 1, \ldots, K} {\delta_p}$;} %
	\STATE{Evaluate $c^{(i)}_{\eta_\phi} = \phi$;}
	\STATE{Invalidate the $\phi$-th line and $\eta_\phi$-th column of $\colk{\bf\Gamma}^{(i)}$;} %
\ENDFOR
\end{algorithmic}
Output: ${\bf c}^{(i)}$.
\end{algorithm}
\normalsize

\section{Power Control Algorithm}\label{sec:PowC}
\colk{We assume a decentralized power allocation procedure, similarly to pilot assignment}. With the purpose of serving the users with a target SINR, the target-SIR-tracking algorithm measures the interference seen by each user, and  assigns to him the exact power to reach the target SINR, unless if this power exceeds the maximum power available. In this case, the maximum power is allocated for this user. It is shown in \cite{Rasti11} that this power allocation procedure is not the most suitable, since assigning the maximum power for the users with poor channel conditions causes an excessive interference for the other users, and waste energy because this user may remain with a low SINR. Being $\hat{\varsigma}^{\rm d}_{k' \ell}$ the target downlink SINR for the $k'$-th user of the $\ell$-th cell, and $\overline{\rho}^{\rm d}_{\ell k'}$ the maximum DL transmit power that can be assigned to him, the target-SIR-tracking algorithm updates power at the $i$-th iteration according to
\begin{equation}\label{eq:TPC_pow}
{\rho}^{\rm d}_{\ell k'}(i) = \min \left[ \hat{\varsigma}^{\rm d}_{k' \ell} \mathcal{I}^{\rm d}_{k' \ell}(i), \overline{\rho}^{\rm d}_{\ell k'} \right],
\end{equation}
in which $\mathcal{I}^{\rm d}_{k' \ell}(i) = {\rho}^{\rm d}_{\ell k'}(i-1)/\varsigma^{\rm d}_{k' \ell}(i-1)$ is the interference plus noise seen by this user (denominator of \eqref{eq:DL_SINR}) divided by $\beta^2_{\ell k' \ell}/\alpha^2_{\ell k'}$. By contrast, the power control algorithm proposed in \cite{Rasti11} updates users' powers as
\begin{eqnarray}\label{eq:OPC_pow}
{\rho}^{\rm d}_{\ell k'}(i) = \left\{\begin{matrix}
\hat{\varsigma}^{\rm d}_{k' \ell} \mathcal{I}^{\rm d}_{k' \ell}(i) & {\rm if} \quad \mathcal{I}^{\rm d}_{k' \ell}(i)\leq \frac{\overline{\rho}^{\rm d}_{\ell k'}}{\hat{\varsigma}^{\rm d}_{k' \ell}},\\ 
\frac{\left(\overline{\rho}^{\rm d}_{\ell k'}\right)^2}{\hat{\varsigma}^{\rm d}_{k' \ell} \mathcal{I}^{\rm d}_{k' \ell}(i)} & {\rm otherwise.}
	\end{matrix}\right.
\end{eqnarray}

One can see that the method of \cite{Rasti11} assigns power to the users in the same way as the target-SIR-tracking algorithm if the target SINR can be achieved for the user at that iteration. Otherwise, instead of allocating him the maximum power, \emph{it allocates him a transmit power inversely proportional to that required to achieve the target}. Thus, besides of saving energy relative to users that cannot reach the target SINR, the interference irradiated to other users also decreases. \colk{This procedure is repeated for a predefined number of iterations $N_{it}$, and the power coefficients for each user can be initialized with half of the maximum transmit power for him, as described in Algorithm \ref{alg:PC_proced}. The convergence of such algorithm is proved in \cite{Rasti11}, and in our simulations we have noted that 10 iterations were sufficient for achieving convergence}. Equivalently, this same procedure can be applied \colk{in UL user} power allocation as
\begin{eqnarray}\label{eq:OPC_pow_UL}
{\rho}^{\rm u}_{k' \ell}(i) = \left\{\begin{matrix}
\hat{\varsigma}^{\rm u}_{k' \ell} \mathcal{I}^{\rm u}_{k' \ell}(i) & {\rm if} \quad \mathcal{I}^{\rm u}_{k' \ell}(i)\leq \frac{\overline{\rho}^{\rm u}_{k' \ell}}{\hat{\varsigma}^{\rm u}_{k' \ell}},\\ 
\frac{\left(\overline{\rho}^{\rm u}_{k' \ell}\right)^2}{\hat{\varsigma}^{\rm u}_{k' \ell} \mathcal{I}^{\rm u}_{k' \ell}(i)} & {\rm otherwise,}
	\end{matrix}\right.
\end{eqnarray}
in which $\hat{\varsigma}^{\rm u}_{k \ell}$ is the target UL SINR of the user, $\overline{\rho}^{\rm u}_{k \ell}$ its maximum UL transmit power, and \small$$\mathcal{I}^{\rm u}_{k' \ell}(i) = \frac{\sum^{L}_{\substack{l = 1 \\ l \neq \ell}} \rho^{\rm u}_{k'l}(i-1) \beta^2_{\ell k' l} + \frac{\alpha^2_{\ell k'}}{\colk{N}} \left(\sum_l \sum_k \rho^{\rm u}_{k l}(i-1) \beta_{\ell k l} + \sigma^2_n \right)}{\beta^2_{\ell k' \ell}}.$$\normalsize

When deploying these power control algorithms, the target SINR parameter should be carefully chosen. If a somewhat lower value is adopted, the algorithm saves energy by delivering just the target SINR to the users, taking low advantage of the resources and providing poor performance for the system. If an excessive target SINR is considered, many poor located users will have their powers gradually turned off in order to provide the desired performance for the other users. Hence, in this paper, the target SINR was chosen in each scenario by finding the value that achieves the higher throughput for 95\% of the users, by means of numerical simulations, as indicated in Table \ref{tab:SINR_tgt}.

\begin{algorithm}
\caption{\small{\colk{Power Control Algorithm}}}
\begin{flushleft}
Input: {$\beta_{j k l}$ and $\overline{\rho}^{\rm d}_{l k}$,  $\forall j,l =\nolinebreak 1,2,\ldots L$, $k= 1,2, \ldots K$, $\sigma_n^2$, $\rho_{\rm p}$.}
\end{flushleft}
\label{alg:PC_proced}
\begin{algorithmic}[1]
\STATE{Initialize ${\rho}^{\rm d}_{j k'}(0) = 0.5 \,\,  \overline{\rho}^{\rm d}_{j k'}$;}
\FOR{$i = 1, 2, \ldots, N_{it}$}
	\FOR{$k' = 1, 2, \ldots, K$}
		\STATE{Evaluate $\mathcal{I}^{\rm d}_{k' j}(i) = {\rho}^{\rm d}_{j k'}(i-1)/\varsigma^{\rm d}_{k' j}(i-1)$;}
		\IF{$\mathcal{I}^{\rm d}_{k' j}(i)\leq \frac{\overline{\rho}^{\rm d}_{j k'}}{\hat{\varsigma}^{\rm d}_{k' j}}$}
			\STATE{Evaluate ${\rho}^{\rm d}_{j k'}(i) = \hat{\varsigma}^{\rm d}_{k' j} \mathcal{I}^{\rm d}_{k' j}(i)$;}
		\ELSE
			\STATE{Evaluate ${\rho}^{\rm d}_{j k'}(i) = \frac{\left(\overline{\rho}^{\rm d}_{j k'}\right)^2}{\hat{\varsigma}^{\rm d}_{k' j} \mathcal{I}^{\rm d}_{k' j}(i)}$;}
		\ENDIF
	\ENDFOR	
\ENDFOR
\end{algorithmic}
Output: ${\rho}^{\rm d}_{j k'}$.
\end{algorithm}
\normalsize

\section{Numerical Results}\label{sec:results}
The Ma-MIMO system performance is numerically evaluated in this Section. We consider $L$ = 7 interfering hexagonal cells of radius 1000m, where $K$ users are uniformly distributed, except in a circle of 100m radius around the cell centered BS \cite{Marinello2016}, with universal frequency and pilot reuse. This is equivalent to say that we consider the performance of a given cell with the interference from 6 nearest-neighbor cells, since only the performance metrics of users in central cell are computed. We consider a system bandwidth of 20MHz, a signal-to-noise ratio of 10dB for pilot, UL data, and DL data transmission, and a TDD architecture similar to \cite{Debbah16}, with $\mathcal{S} = 100$\footnote{Which can represent, for instance, a scenario with 1 ms of channel coherence time and 100 KHz of coherence bandwidth.} symbols, of which $K$ are dedicated to UL pilot transmissions, and the remainder are equally divided between UL and DL data transmissions ($\xi^{\rm u} = \xi^{\rm d} = 0.5$). The path loss decay exponent was adopted as 3.8, and the standard deviation of the log-normal shadowing was assumed to be 8dB. \colk{Besides, Table \ref{tab:parameters} describes the notation adopted when referring to the different techniques investigated in this Section.}

\begin{table*}[!htbp]
\caption{\colk{Acronyms for the investigated schemes.}}
\vspace{-2mm}
\centering
\small
\colk{
{\renewcommand{\arraystretch}{1.4}%
\begin{tabular}{rl}
\hline
\bf Acronym & \bf Technique\\
\hline
MaxSINR DL & The PA method that obtains the highest mean DL SINR via exhaustive search \\
MaxMinSINR DL & The PA method that solves \eqref{eq:MaxminSINR_DL_PA} via exhaustive search \\
H-MaxMinSINR DL & Our proposed PA method that solves \eqref{eq:MaxminSINR_DL_PA} applying Algorithm \ref{alg:PA_proced} \\
H-MaxMinSINR DL PA+PC & H-MaxMinSINR DL combined with the power control of Algorithm \ref{alg:PC_proced} \\
MaxSINR UL & The PA method that obtains the highest mean UL SINR via exhaustive search \\
MaxMinSINR UL & The PA method that solves \eqref{eq:MaxminSINR_UL_PA} via exhaustive search \\
H-MaxMinSINR UL & The PA method that solves \eqref{eq:MaxminSINR_UL_PA} proposed in \cite{SPA_Zhu15} \\
H-MaxMinSINR UL PA+PC & H-MaxMinSINR UL combined with the power control of Algorithm \ref{alg:PC_proced} \\
MaxTC & The PA method that obtains the highest mean Total Capacity via exhaustive search \\
MaxMinTC & The PA method that solves \eqref{eq:MaxminCtt_PA} via exhaustive search \\
H-MaxMinTC & Our proposed PA method that solves \eqref{eq:MaxminCtt_PA} applying Algorithm \ref{alg:PA_proced}\\
H-MaxMinTC PA+PC & H-MaxMinTC combined with the power control of Algorithm \ref{alg:PC_proced}\\
\hline
\end{tabular}}}
\label{tab:parameters}
\end{table*}
\normalsize

\subsection{\colk{Performance Comparison between Max-Min and Average PA Optimization Approaches}}\label{sec:perf_comp}

\colk{Our first objective in this Section is to justify the max-min approach choice when solving the PA optimization problem. For this purpose, we compare the performance of max-min approach with the conventional average approach, \emph{i.e.}, a PA strategy that finds the highest average SINR performance via exhaustive search. As described in Table \ref{tab:parameters}, we refer to the PA method that obtains the highest mean DL SINR as MaxSINR DL, the one that obtains the highest mean UL SINR as MaxSINR UL, and the one that obtains the highest mean total capacity as MaxTC. Besides, we refer to the PA that solves \eqref{eq:MaxminSINR_UL_PA} as MaxminSINR UL, the one solving \eqref{eq:MaxminSINR_DL_PA} as MaxminSINR DL, and the solution of \eqref{eq:MaxminCtt_PA} as MaxminTC. Considering $N = 128$ antennas and $K = 4$ users, Table \ref{tab:mean_rates} shows the performance achieved by the different PA schemes in terms of mean achievable rates. As expected, MaxSINR DL, MaxSINR UL and MaxTC obtain the best mean performance on their respective metric (shown in bold in the Table). On the other hand, MaxminSINR DL, MaxminSINR UL, and MaxminTC presented a small decrease in mean achievable rates when compared to that obtained by MaxSINR DL, MaxSINR UL and MaxTC.}

\footnotesize
\begin{table*}[!htbp]
\caption{\colk{Mean Achievable Rates for $N=128$ and $K=4$.}}
\vspace{-2mm}
\centering
\small
\colk{
{\renewcommand{\arraystretch}{1.4}%
\begin{tabular}{|l|l|l|l|}
\hline
\bf PA Technique & \bf DL& \bf UL& \bf TC\\
\hline
\hline
Random & 28.46Mbps & 24.79Mbps & 53.25Mbps \\
 \hline
MaxMinSINR DL & 30.24Mbps & 24.35Mbps & 54.59Mbps \\
 \hline
MaxSINR DL & {\bf 30.36}Mbps & 24.45Mbps & 54.8Mbps \\
 \hline
MaxMinSINR UL & 29.13Mbps & 24.22Mbps & 53.35Mbps \\
 \hline
MaxSINR UL & 28.07Mbps & {\bf 25.48}Mbps & 53.55Mbps \\
 \hline
MaxMinTC & 29.79Mbps & 24.41Mbps & 54.21Mbps \\
 \hline
MaxTC & 29.92Mbps & 25.32Mbps & {\bf 55.53}Mbps \\
 \hline
\end{tabular}}}
\label{tab:mean_rates}
\end{table*}
\normalsize

\begin{figure}[!htbp]
\centering
\includegraphics[width=0.95\columnwidth]{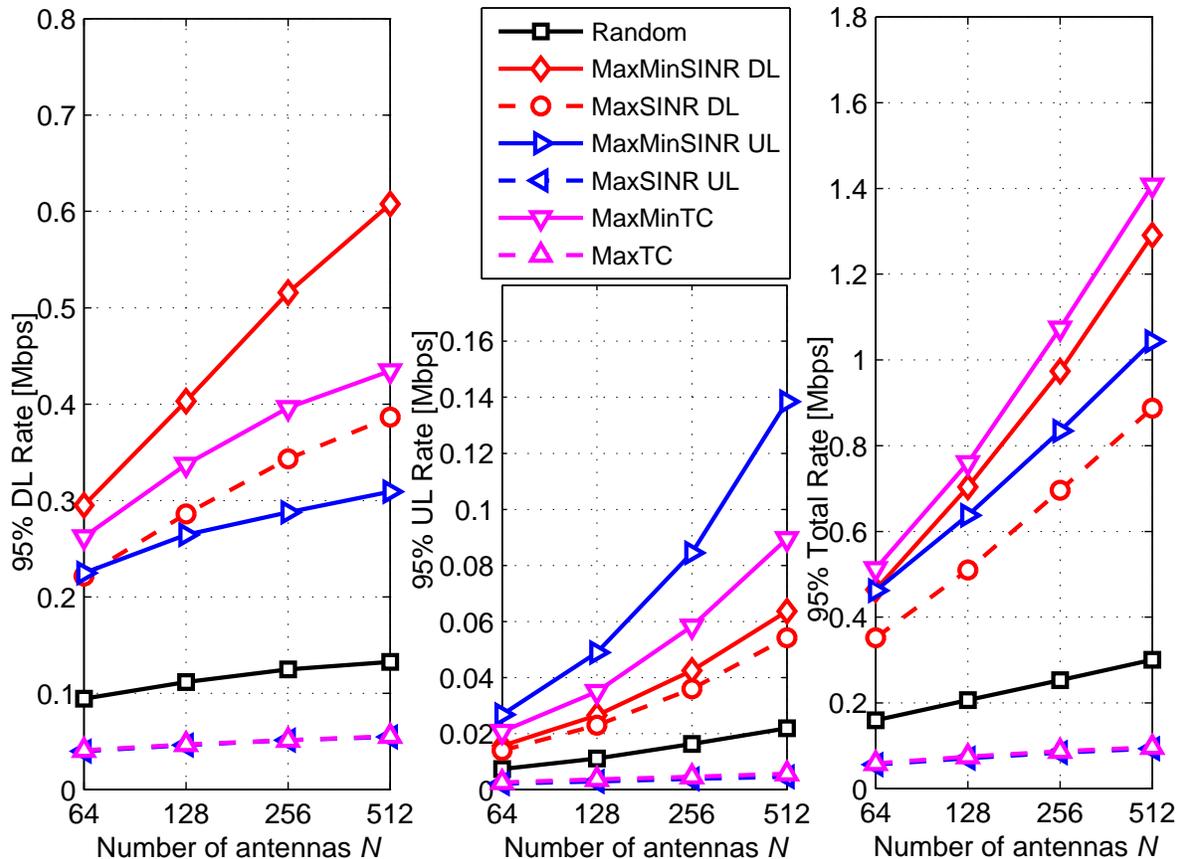}
\vspace{-4mm}
\caption{\colk{95\%-likely Rate for $K=4$: \colk{a)} DL; \colk{b)} UL; \colk{c)} Total.}}
\label{fig:Rates_5pc_AvxMm_K4}
\end{figure}

\begin{figure}[!htbp]
\centering
\includegraphics[width=0.95\columnwidth]{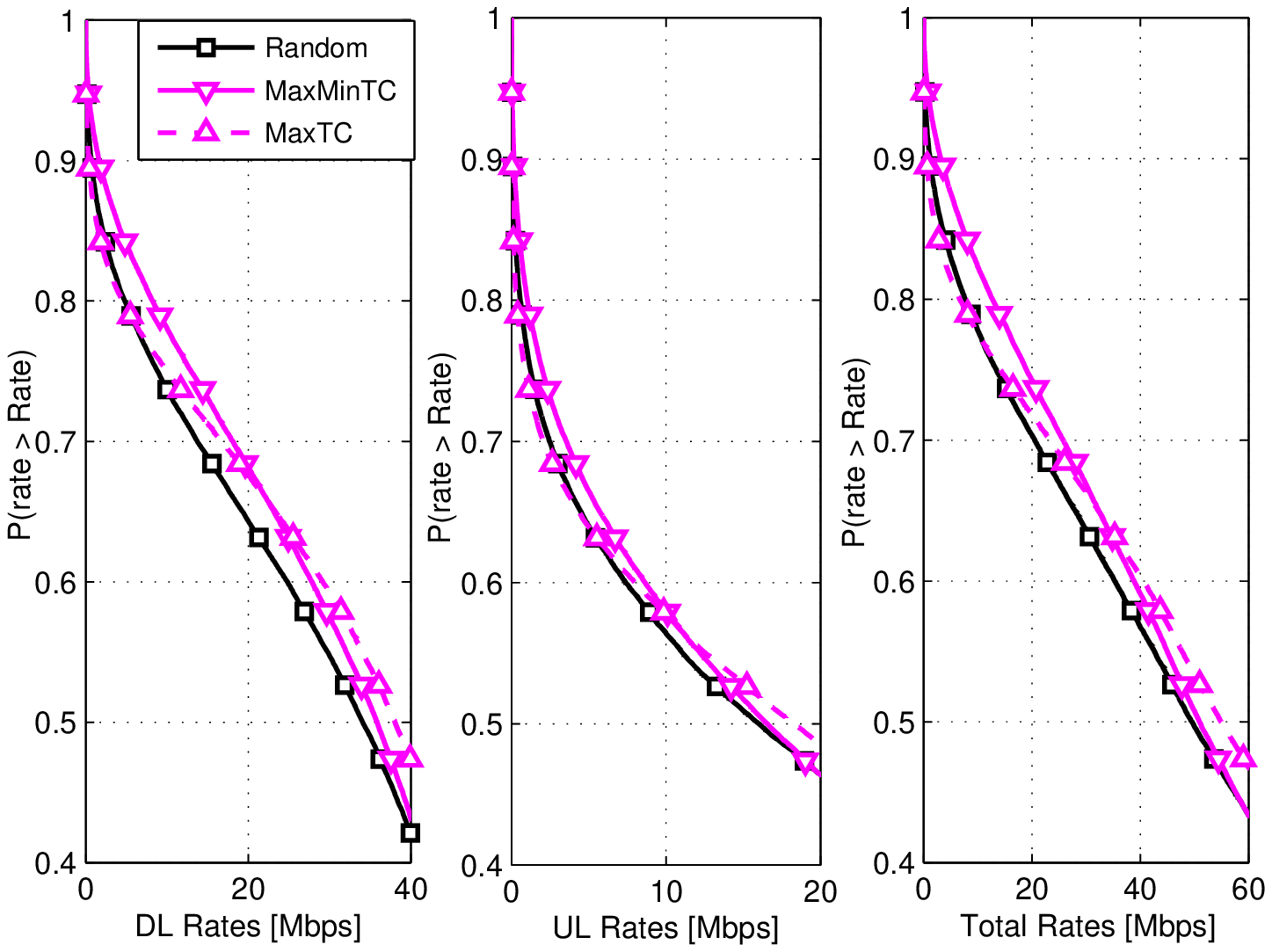}
\vspace{-4mm}
\caption{\colk{Fraction of users above a given Rate for $N = 128$ and $K=4$: \colk{a)} DL; \colk{b)} UL; \colk{c)} Total.}}
\label{fig:Rates_N128_AvxMm_K4}
\end{figure}

\colk{The main advantage of adopting the max-min approach instead of the average approach when solving the PA problem can be shown in Figure \ref{fig:Rates_5pc_AvxMm_K4}, which compares the 95\%-likely rate achieved by each PA technique with the increasing number of antennas. The 95\%-likely rate corresponds to the rate assured to the users with probability of 95\%, and thus can be used to compare how fair is the performance of a given technique. As one can note from Figure \ref{fig:Rates_5pc_AvxMm_K4}, a significant gain can be seen when comparing a max-min based PA technique with its average based counterpart in terms of its respective performance metric. For example with 128 BS antennas, there is a 0.12Mbps improvement in the 95\%-likely DL rate of MaxminSINR DL in comparison with MaxSINR DL. Similarly, there is 0.05Mbps improvement in the 95\%-likely UL rate of MaxminSINR UL in comparison with MaxSINR UL, and a 1.01Mbps improvement in the 95\%-likely total capacity of MaxminTC in comparison with MaxTC. The reason why a higher 95\%-likely performance is achieved at the same time that a lower average performance when adopting the max-min approach instead of the average one is depicted in Figure \ref{fig:Rates_N128_AvxMm_K4} for the case of total capacity metric. A higher average performance is obtained by MaxTC PA providing higher rates for the users in the best channel conditions, while providing poor rates to the users in the worst channel conditions. On the other hand, MaxminTC PA aims to provide the best possible performance to the users in the worst channel conditions, while the users with higher channel coefficients still present good performances. This behavior is characterized by the PA performance curves crossing each other in Figure \ref{fig:Rates_N128_AvxMm_K4}, which means that the portion of users with very poor performances is higher in MaxTC than in MaxminTC, while the portion of users with very high performances is lower in MaxminTC than in MaxTC.}

\vspace{-1mm}

\vspace{-1mm}

\subsection{\colk{Performance of Heuristic Approach}}

\begin{figure}[!htbp]
\centering
\includegraphics[width=0.95\columnwidth]{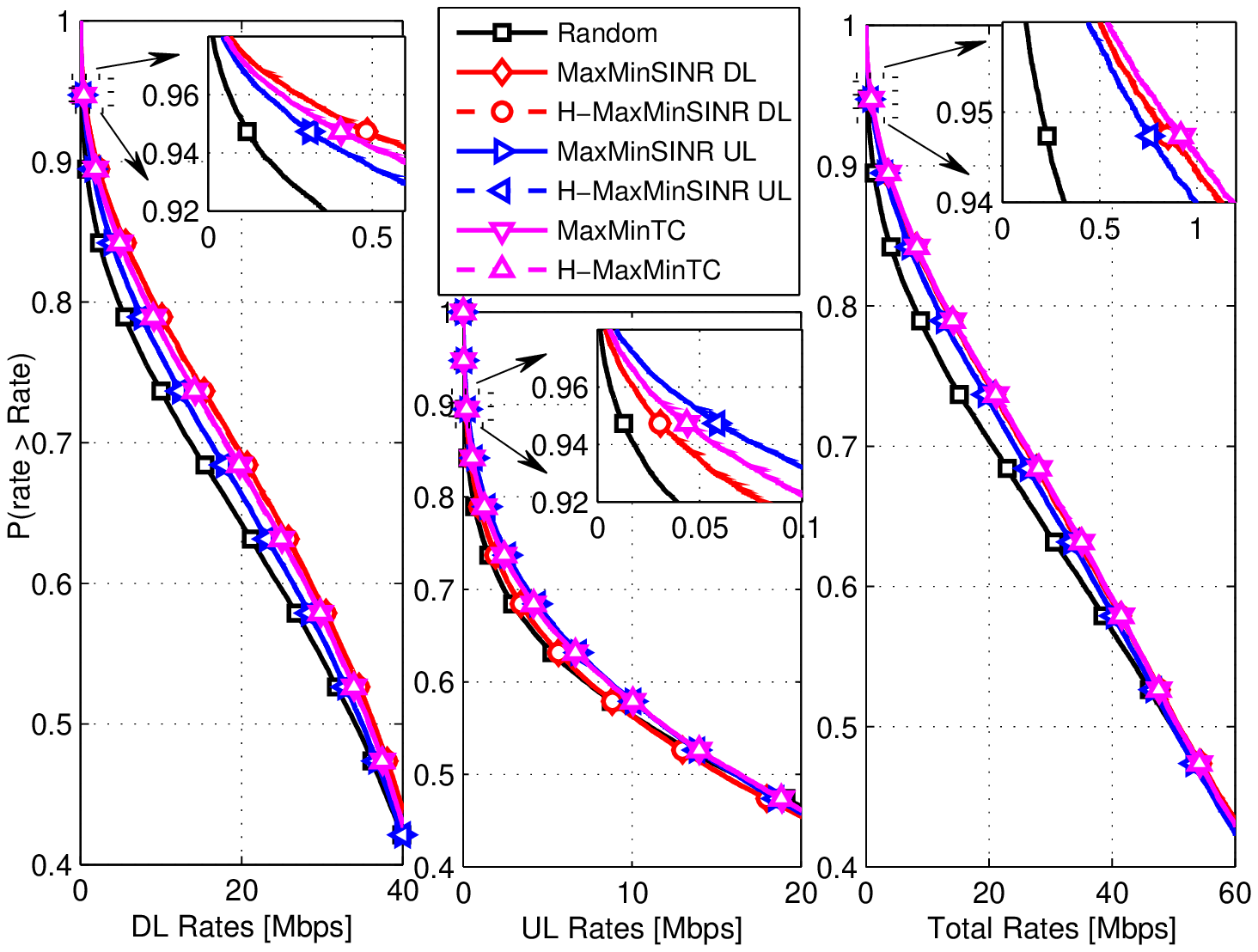}
\vspace{-4mm}
\caption{Fraction of users above a given Rate for $N = 128$ and $K=4$: \colk{a)} DL; \colk{b)} UL; \colk{c)} Total.}
\label{fig:Rates_N128_K4}
\end{figure}

\begin{figure}[!htbp]
\centering
\includegraphics[width=0.95\columnwidth]{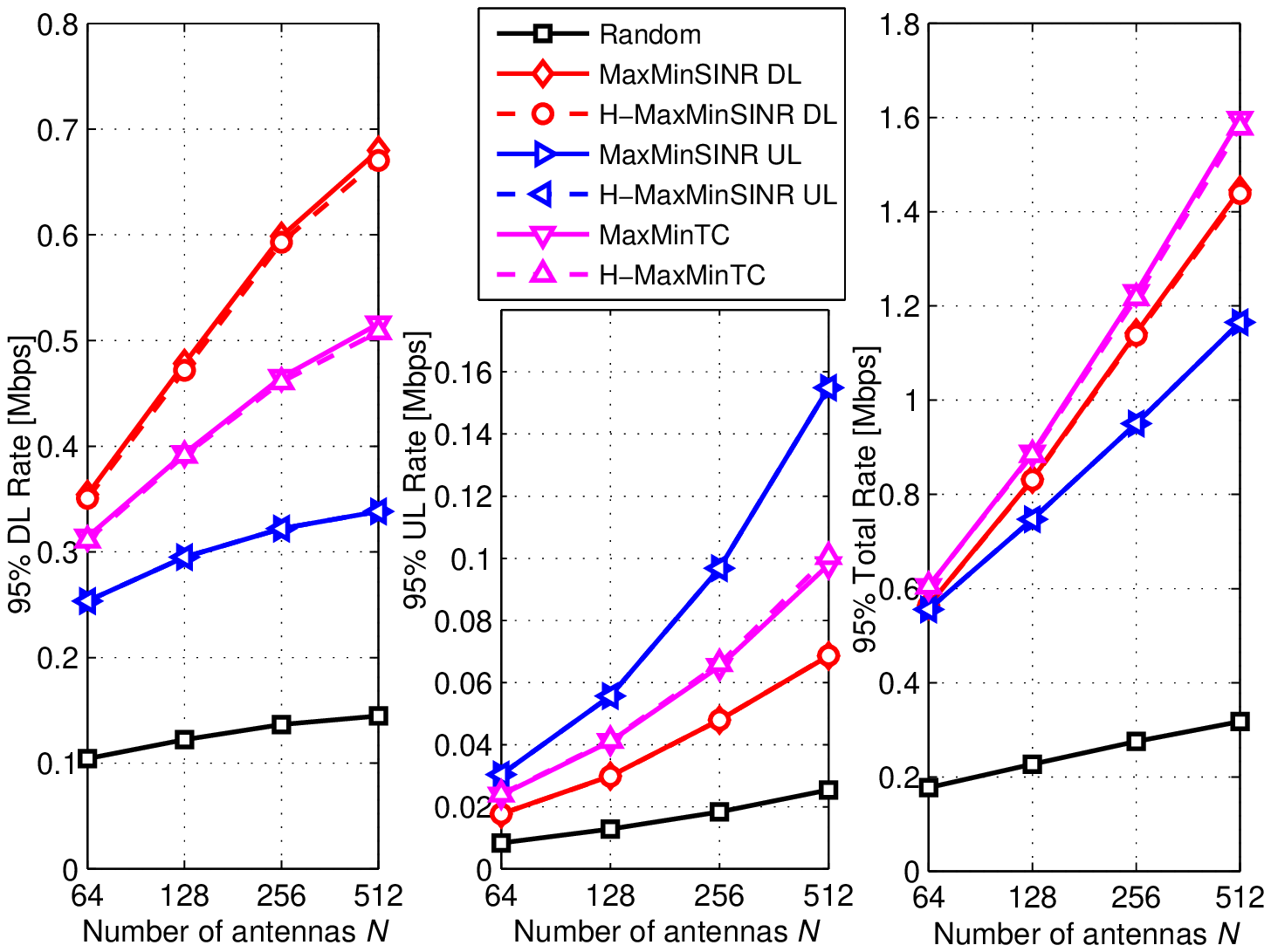}
\vspace{-4mm}
\caption{95\%-likely Rate for $K=4$: \colk{a)} DL; \colk{b)} UL; \colk{c)} Total.}
\label{fig:95PC_Rates_K4}
\end{figure}

\colk{We analyze in this subsection the performance obtained by the near optimal low-complexity PA solution of equations \eqref{eq:MaxminSINR_UL_PA}, \eqref{eq:MaxminSINR_DL_PA} and \eqref{eq:MaxminCtt_PA}. Besides, we discard the average based PA techniques like MaxSINR DL, MaxSINR UL, and MaxTC, since our main objective is to improve the performance guaranteed for the majority of users.} The heuristic approach of each solution is referred by the respective method preceded by "H". \colk{Our objective herein} is to demonstrate that the \emph{heuristic solutions} of MaxminSINR UL, MaxminSINR DL, and MaxminTC \emph{achieve practically the same performance of that attained with exhaustive search}, but with a feasible complexity. For this sake, Figure \ref{fig:Rates_N128_K4} depicts the fraction of users above a given rate for 128 BS antennas, in terms of DL, UL, and total rate, and Figure \ref{fig:95PC_Rates_K4} shows the 95\%-likely rate achieved by each technique with increasing number of antennas. In both cases, an uniform power allocation policy was assumed for simplicity, and $K = 4$ users in order to allow the evaluation of exhaustive search in a feasible time. It is noteworthy from Figure \ref{fig:Rates_N128_K4} the very unfair behavior of primitive Ma-MIMO systems (with no pilot and power allocation policies), which provides very high data rates for some portion of the users, while providing low quality of service for others. One can note also that appreciable improvements can be achieved with PA techniques, \colk{represented in Figure \ref{fig:Rates_N128_K4} by a slight horizontal shift of the PA curves in the region of high probability}. The conflicting behavior of UL and DL optimization metrics is also represented in the Figure, since while MaxminSINR DL achieves the best performance in DL, its UL performance is close to that attained by random PA. The opposite occurs with MaxminSINR UL. On the other hand, MaxminTC achieves good performance in both directions, at the same time that it achieves the best performance in terms of total rates. Besides, no significant performance loss can be seen for any heuristic PA when compared with its respective exhaustive search solution. Similar findings can be taken from Figure \ref{fig:95PC_Rates_K4}, that also reveals that is always beneficial improving the number of antennas in BS. \colk{One can note from Figure \ref{fig:95PC_Rates_K4} that MaxminSINR DL achieves the best 95\%-likely DL rate, while MaxminSINR UL achieves the best 95\%-likely UL rate and MaxminTC achieves the best 95\%-likely total capacity. Besides, its heuristic counterparts produce almost the same 95\%-likely performance of them.}

\vspace{-1mm}

\vspace{-1mm}

\subsection{\colk{Performance of Pilot Assignment with Power Control}}

\begin{figure}[!htbp]
\centering
\includegraphics[width=0.95\columnwidth]{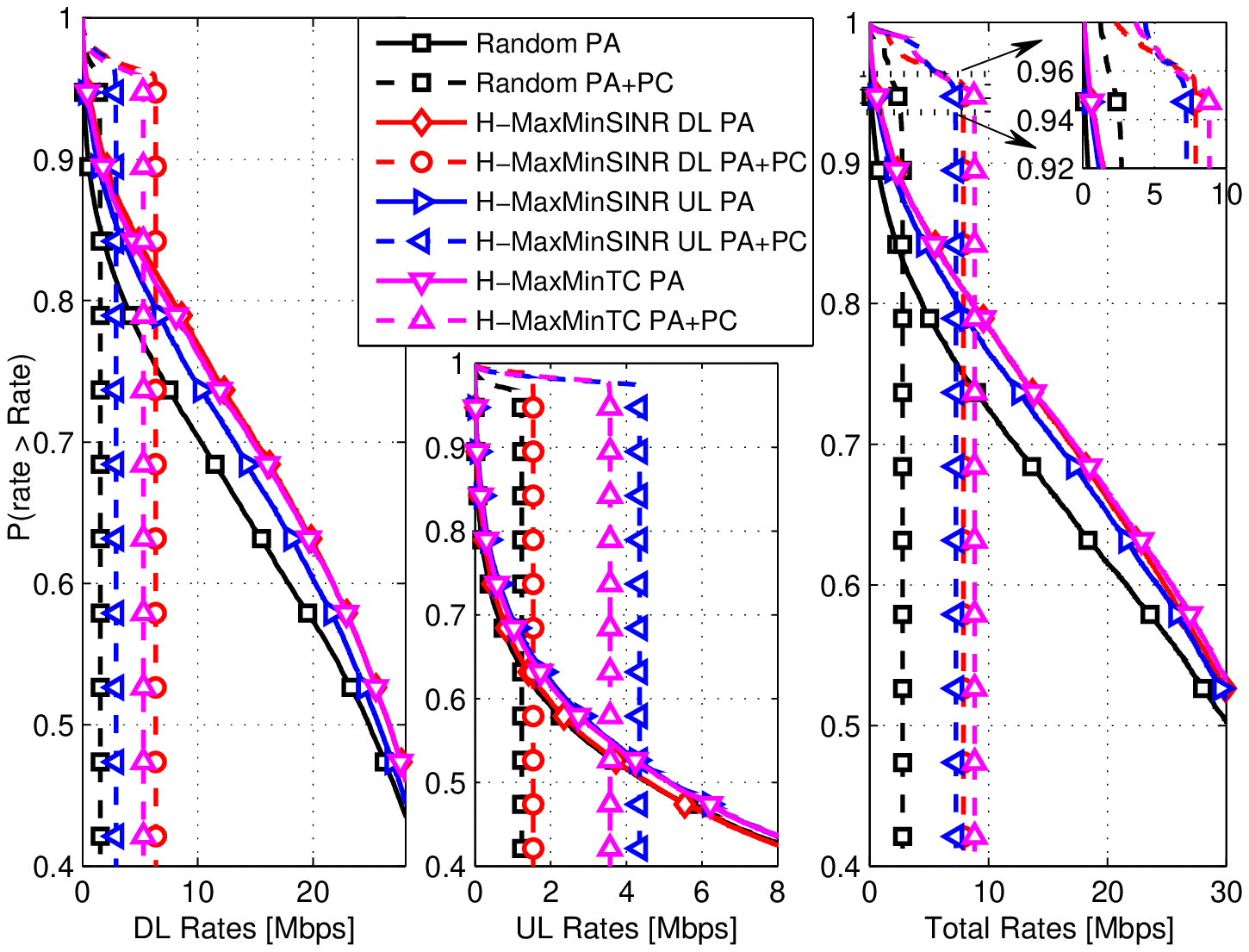}
\vspace{-4mm}
\caption{Fraction of users above a given Rate for $N = 128$ and $K=10$: \colk{a)} DL; \colk{b)} UL; \colk{c)} Total. \colk{Power control evaluated with target SINR's of Table \ref{tab:SINR_tgt}}.}
\label{fig:Rates_N128_K10_PC}
\end{figure}

\begin{figure}[!htbp]
\centering
\includegraphics[width=0.95\columnwidth]{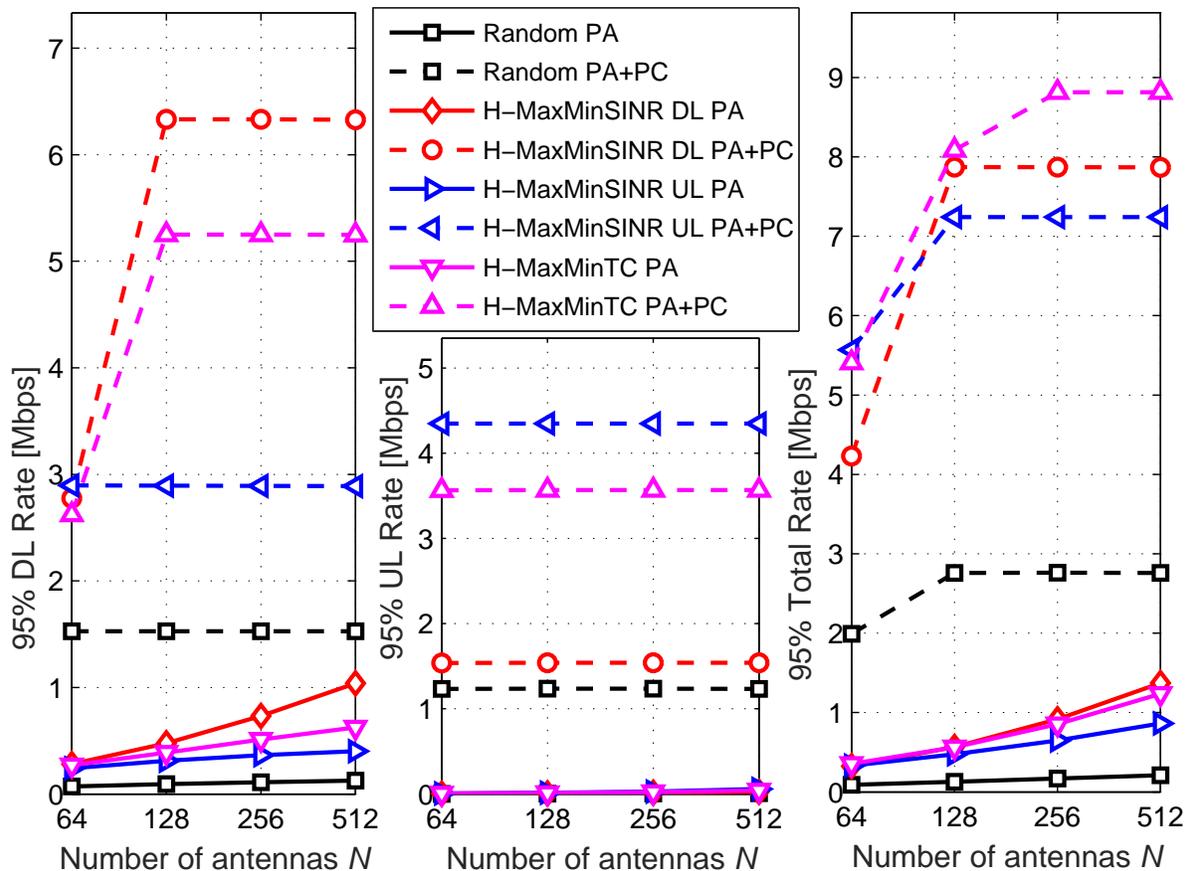}
\vspace{-4mm}
\caption{95\%-likely Rate for $K=10$: \colk{a)} DL; \colk{b)} UL; \colk{c)} Total. \colk{Power control evaluated with target SINR's of Table \ref{tab:SINR_tgt}}.}
\label{fig:95PC_Rates_K10}
\end{figure}

Having demonstrated the good performances achieved by the heuristic PA techniques, our goal now is to demonstrate that the power control algorithm achieves a much improved performance when combined with an appropriate PA scheme. Besides, we discard the exhaustive search PA algorithms in order to enable a higher number of users. For $K=10$ and employing power control, Figure \ref{fig:Rates_N128_K10_PC} shows the fraction of users above a given rate for $N=128$, while Figure \ref{fig:95PC_Rates_K10} shows the 95\%-likely rate achieved with increasing number of antennas. The target SINR's performances of power control algorithm for each scheme were empiricaly chosen in order to provide the higher throughput for 95\% of the users, as indicated in Table \ref{tab:SINR_tgt}. It can be seen that much more uniform user performances are achieved with the power control \colk{scheme of Algorithm \ref{alg:PC_proced}}, and that the assured performance can be considerably improved with a proper choice of the PA employed. For example, as can be seen in Figure \ref{fig:95PC_Rates_K10}, with random assignment the 95\%-likely DL rate increases from 92.5 kbps to 1.528 Mbps with $N=128$ antennas when employing power control, while this DL rate increase is from 388.3 kbps to 5.251 Mbps when the H-MaxminTC is the PA adopted. Similarly, in the UL, the 95\%-likely UL rate with random assignment increases from 3.4 kbps to 1.235 Mbps when employing power control, and from 11.4 kbps to 3.566Mbps with H-MaxminTC as PA policy and $N=128$. It can also be seen that H-MaxminTC with power control has the ability of jointly assure an appreciable quality of service in both DL and UL, contrary to H-MaxminSINR DL and H-MaxminSINR UL that assure good rates only on its preferential direction. \colk{These results demonstrate that much more significant performance improvements can be obtained from the power control algorithm with a proper choice of the PA method.}

\vspace{-1mm}

\small
\begin{table*}[!htbp]
\caption{Target SINR's of power control algorithm adopted for each PA technique.}
\vspace{2mm}
\centering
\footnotesize
\begin{tabular}{|c|c|c|c|}
\hline
 \bf Number of users	 & \bf PA Scheme    & \bf DL  & \bf UL \\
\hline
\hline
\multirow{4}{*}{\large{$K$ = 10}} & Random & -9 dB & -10 dB \\\cline{2-4}
 & H-MaxminSINR DL & -2 dB & -9 dB \\\cline{2-4}
 & H-MaxminSINR UL & -6 dB & -4 dB \\\cline{2-4}
 & H-MaxminTC & -3 dB & -5 dB \\
\hline
\hline
\multirow{4}{*}{\large{$K$ = 32}} & Random & -11 dB & -14 dB \\\cline{2-4}
 & H-MaxminSINR DL & -7 dB & -11 dB \\\cline{2-4}
 & H-MaxminSINR UL & -9 dB & -7 dB \\\cline{2-4}
 & H-MaxminTC & -7 dB & -8 dB\\
\hline
\end{tabular}
\label{tab:SINR_tgt}
\end{table*}
\normalsize

Figures \ref{fig:Rates_N128_K32_PC} and \ref{fig:95PC_Rates_K32} do the same with $K = 32$ users, from which similar findings can be taken. While the 95\%-likely DL rate for $N=128$ antennas under random pilot assignment increase from 41.3 kbps to 746.2 kbps, this DL rate increase with H-MaxminTC is from 191.6 kbps to 1.782 Mbps. Similarly for the 95\%-likely UL rate with $N=128$, this gain under random PA is from 0.5 kbps to 382.5 kbps, and from 2.0 kbps to 1.442 Mbps with H-MaxminTC PA. The decrease in the assured rates with this higher number of users is due not only to the increased multiuser interference, but also to the increased pilot overhead, necessary to obtain CSI estimates of this increased number of users. On the other hand, it allows the PA schemes to achieve improved gains in comparison with random assignment because of the greater multiuser diversity.

\vspace{-1mm}

\subsection{\colk{Performance with Target Rates}}

\begin{figure}[!htbp]
\centering
\includegraphics[width=0.95\columnwidth]{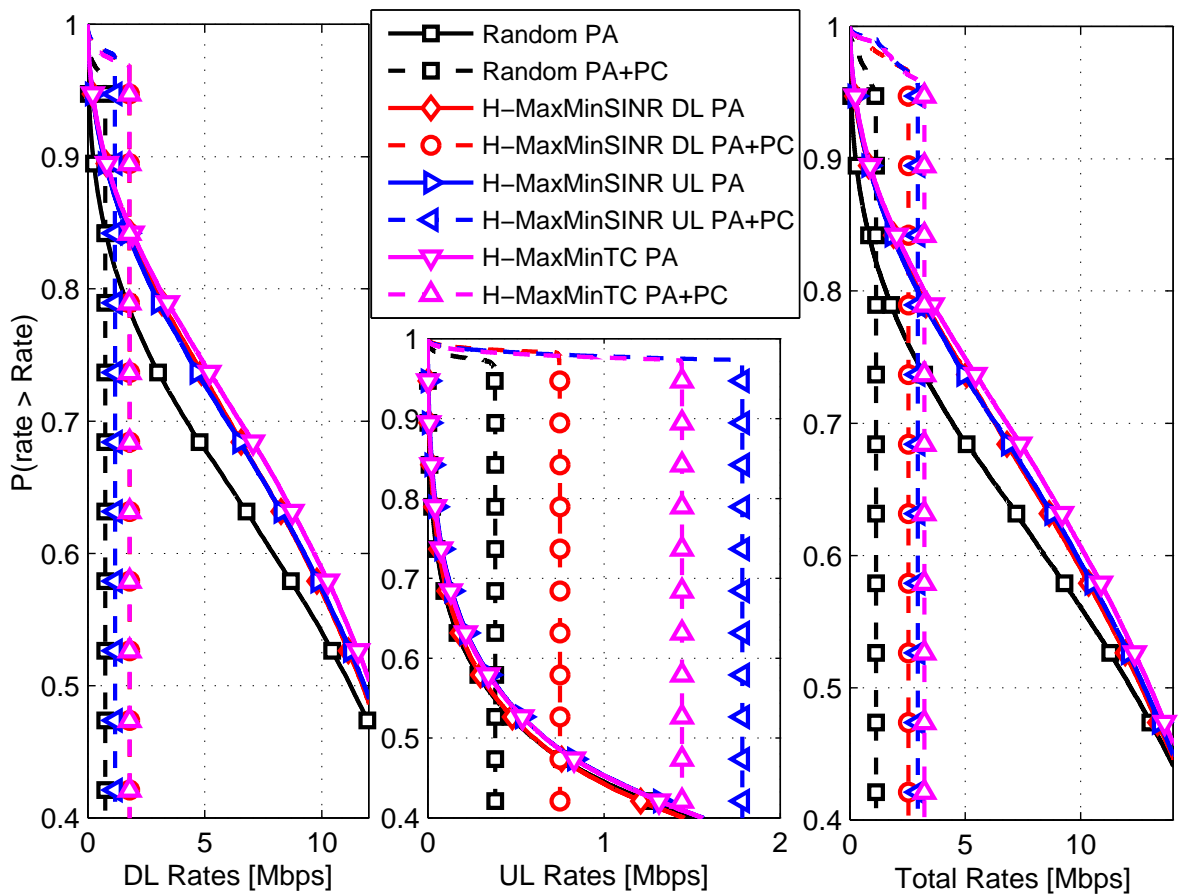}
\vspace{-4mm}
\caption{Fraction of users above a given Rate for $N = 128$ and $K=32$: \colk{a)} DL; \colk{b)} UL; \colk{c)} Total. \colk{Power control evaluated with target SINR's of Table \ref{tab:SINR_tgt}}.}
\label{fig:Rates_N128_K32_PC}
\end{figure}

\begin{figure}[!htbp]
\centering
\includegraphics[width=0.95\columnwidth]{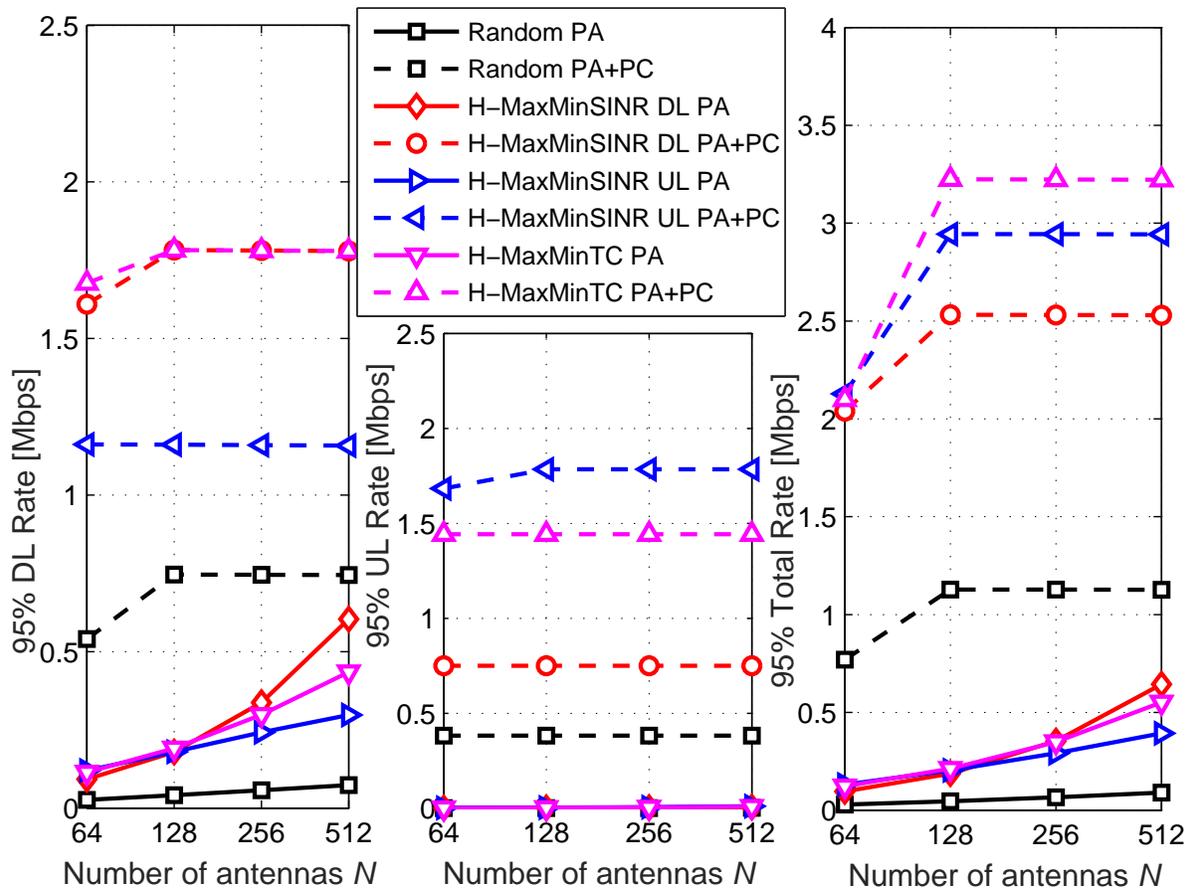}
\vspace{-4mm}
\caption{95\%-likely Rate for $K=32$: \colk{a)} DL; \colk{b)} UL; \colk{c)} Total. \colk{Power control evaluated with target SINR's of Table \ref{tab:SINR_tgt}}.}
\label{fig:95PC_Rates_K32}
\end{figure}

\begin{figure}[!htbp]
\centering
\includegraphics[width=0.95\columnwidth]{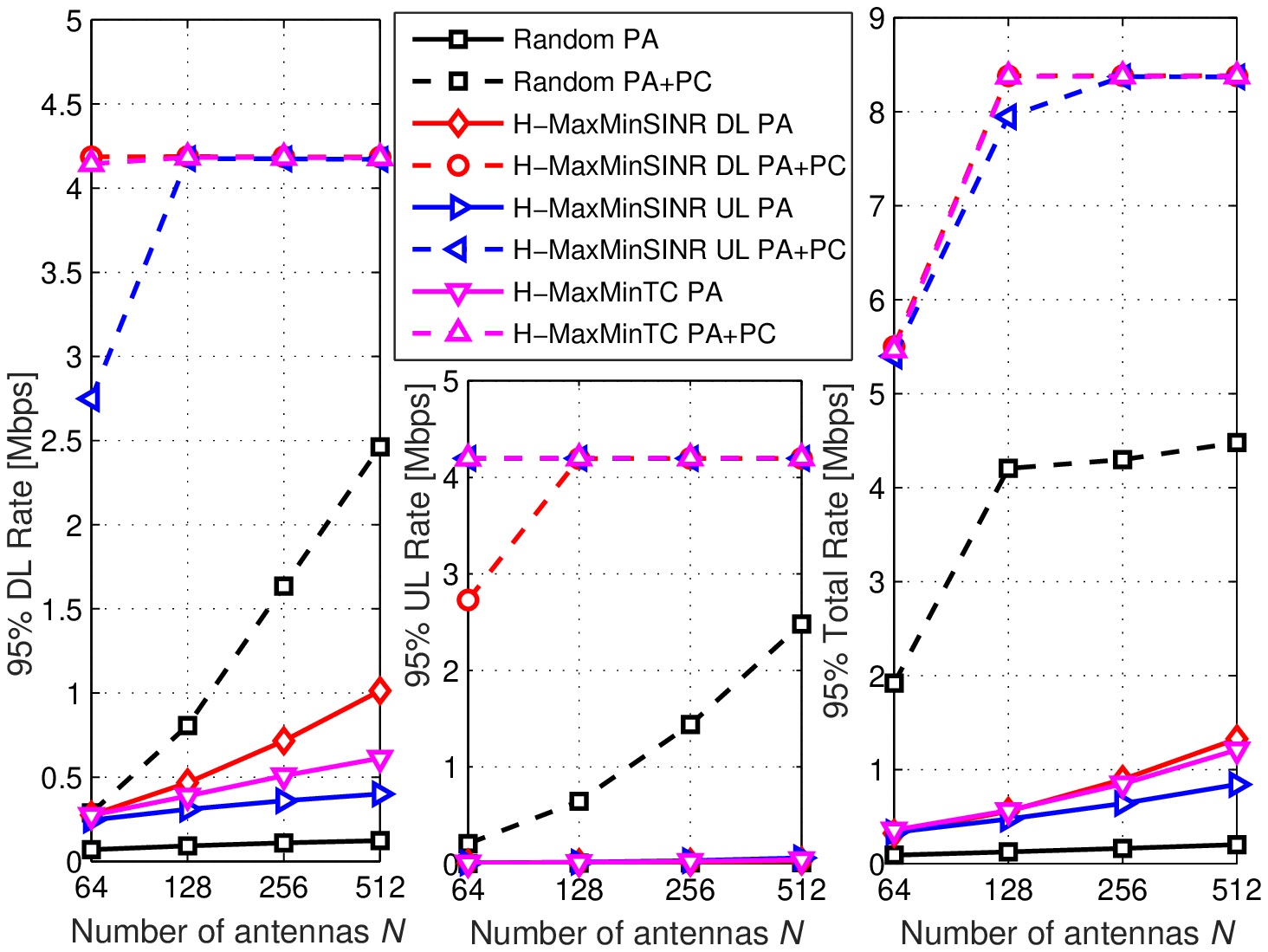}
\vspace{-4mm}
\caption{95\%-likely Rate for $K=10$: \colk{a)} DL; \colk{b)} UL; \colk{c)} Total. \colk{Power control evaluated with a target rate of 4.2 Mbps per user, for both UL and DL}.}
\label{fig:95PC_Rates_K10_CapTg}
\end{figure}

\begin{figure}[!htbp]
\centering
\includegraphics[width=0.95\columnwidth]{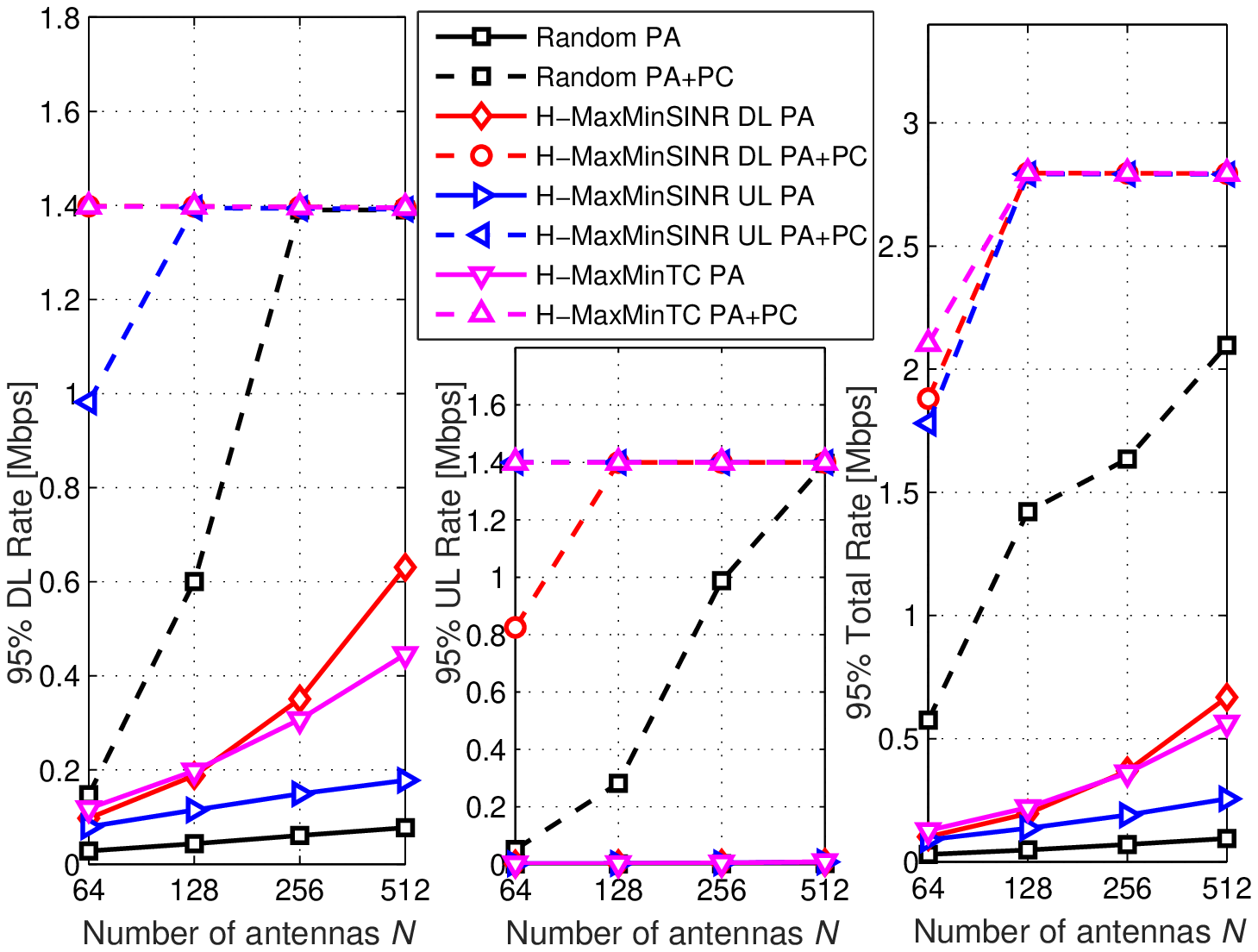}
\vspace{-4mm}
\caption{95\%-likely Rate for $K=32$: \colk{a)} DL; \colk{b)} UL; \colk{c)} Total. \colk{Power control evaluated with a target rate of 1.4 Mbps per user, for both UL and DL}.}
\label{fig:95PC_Rates_K32_CapTg}
\end{figure}

Performance results with power control as depicted in Figures \ref{fig:Rates_N128_K10_PC} to \ref{fig:95PC_Rates_K32} were obtained after a careful but non-exhaustive process of finding suitable target SINR performances for each scheme under the considered scenarios, as summarized in Table \ref{tab:SINR_tgt}. However, this values of target performances are dependent of the instantaneous conditions of the communications systems, like the number of users served and pilot and data SNR's, which are dynamically changing. Thus, a more practical scenario consists in fixing a DL and UL target rates, from which the target SINR's input parameters of the power control algorithm are obtained. Under this strategy, Figure \ref{fig:95PC_Rates_K10_CapTg} shows the 95\%-likely rates obtained when fixing a target \colk{per user} rate of 4.2 Mbps for both DL and UL when serving $K=10$ users, while Figure \ref{fig:95PC_Rates_K32_CapTg} does the same with 1.4 Mbps of target rates for both DL and UL when serving $K=32$ users. It can be seen from Figure \ref{fig:95PC_Rates_K10_CapTg} that H-MaxminTC PA with power control is able to assure the target rates in both DL and UL even with just $N=64$ antennas, \colk{in a situation where the multiuser interference is only partially supressed by the moderate number of BS antennas. On the other hand, H-MaxminSINR DL PA with power control fails in this objective in UL for $N=64$, while H-MaxminSINR UL could not achieve this target performance in DL with this same number of antennas.} It is important to note that even with the 95\%-likely DL and UL rate of 4.2 Mbps assured by H-MaxminTC PA with power control, it is not assured a 95\%-likely total rate of 8.4 Mbps, since the set of users that do not achieve the target DL rate is not necessarily the same that do not achieve the target UL rate. Very similar conclusions can be made from Figure \ref{fig:95PC_Rates_K32_CapTg}, which shows that H-MaxminTC PA with power control is able to assure an appreciable 95\%-likely DL and UL \colk{per user} rates of 1.4 Mbps for $K=32$ users with only $N=64$ BS antennas. \colk{Again, these results demonstrate the great potential of PA techniques in conjunction with power control algorithms in providing improved and homogeneous performance for a larger number of users in both DL and UL.}

\vspace{-1mm}

\section{Discussion and Final Remarks}\label{sec:concl}
The \colk{PA} optimization problem was addressed in this paper. Different from previous works, we have investigated the problem from both UL and DL perspectives. \colk{We demonstrate that the max-min approach when solving the PA problem is much more effective than the average approach when aiming to provide a good performance for the majority of the users.} Our analysis have \colk{also} shown that, due to the different characteristics of UL and DL SINR expressions in Ma-MIMO, the PA optimization problems in UL and DL are conflicting. If the UL performance is optimized, it incurs in a limited performance in DL, and vice-versa. Thus defining a simple alternative metric, {\it i.e.}, the total capacity, one can find a PA strategy for Ma-MIMO that achieves promising performance in both directions simultaneously. To avoid exhaustive search of factorial order, a heuristic solution capable of finding a near-optimal solution expending reduced computational complexity has been also proposed. Finally, we have adapted the target-SIR-tracking power control algorithm to our scenario of massive MIMO systems with finite number of BS antennas, and the investigated PA schemes have been combined with it. Our results demonstrated that much more improved gains can be achieved by this power control algorithm if applied after a suitable pilot assignment procedure. For example, our proposed H-MaxminTC PA scheme with power control was able to provide a 4.2 Mbps \colk{per user} rate for both DL and UL with 95\% probability when serving 10 users with only 64 antennas at BS, while an assured symmetric \colk{per user} rate of 1.4 Mbps is achieved when serving 32 users with 64 antennas. 

\vspace{-1mm}

 \section*{Acknowledgement}
 This work was supported in part by the National Council for Scientific and Technological Development (CNPq) of Brazil under Grant 304066/2015-0 and in part by Londrina State University - Paraná State Government (UEL).

\vspace{-1mm}

\end{document}